\documentclass[reprint,amsmath,amssymb,aps,nobibnotes, superscriptaddress]{revtex4-1}
\usepackage{graphicx}
\usepackage{epstopdf}
\usepackage{dcolumn}
\usepackage{bm}
\usepackage{color}

\begin{document}

\title{The effect of interface roughness on exchange bias in La$_{0.7}$Sr$_{0.3}$MnO$_3$ - BiFeO$_3$ heterostructures}

\author{Mehran Vafaee}
\affiliation{Institut f\"{u}r Physik, Johannes-Gutenberg-Universit\"{a}t of Mainz, Staudingerweg 7, 55128 Mainz, Germany}
\author{Simone Finizio}
\affiliation{Institut f\"{u}r Physik, Johannes-Gutenberg-Universit\"{a}t of Mainz, Staudingerweg 7, 55128 Mainz, Germany}
\author{Hakan Deniz}
\affiliation{Max-Planck Institute of Microstructure Physics, Weinberg 2, 06120 Halle (Saale), Germany}
\author{Dietrich Hesse}
\affiliation{Max-Planck Institute of Microstructure Physics, Weinberg 2, 06120 Halle (Saale), Germany}
\author{Hartmut Zabel}
\affiliation{Institut f\"{u}r Physik, Johannes-Gutenberg-Universit\"{a}t of Mainz, Staudingerweg 7, 55128 Mainz, Germany}
\author{Gerhard Jakob}
\affiliation{Institut f\"{u}r Physik, Johannes-Gutenberg-Universit\"{a}t of Mainz, Staudingerweg 7, 55128 Mainz, Germany}
\author{Mathias Kl\"{a}ui}
\affiliation{Institut f\"{u}r Physik, Johannes-Gutenberg-Universit\"{a}t of Mainz, Staudingerweg 7, 55128 Mainz, Germany}

\date{September 11, 2015}%

\begin{abstract}
  We characterized the interfaces of heterostructures with different stack sequences of La$_{0.7}$Sr$_{0.3}$MnO$_3$/BiFeO$_3$ (LSMO/BFO) and BFO/LSMO using TEM revealing sharp and rough interfaces, respectively. Magnetometry and magnetoresistance measurements do not show a detectable exchange bias coupling for the multistack with sharp interface. Instead, the heterostructures with rough and chemically intermixed interfaces exhibit a sizable exchange bias coupling. Furthermore, we find a temperature-dependent irreversible magnetization behavior and an exponential decay of coercive and exchange bias field with temperature suggesting a possible spin-glass-like state at the interface of both stacks.
\end{abstract}
\maketitle

The electric field control of magnetism in artificial multiferroics consisting of ferroelectric and ferromagnetic layers has been studied extensively in the recent years. In this context, BFO as a natural multiferroic (antiferromagnetic and ferroelectric) has become a canonical compound of interest due to the scarcity of room temperature multiferroics in nature\,\cite{Ramesh:07}. The realization of electric field controlled magnetic devices has been practiced by combining BFO with the ferromagnetic oxide LSMO at whose interface, the ferromagnetic order of LSMO is coupled with the antiferromagnetic order and ferroelectric orders of BFO through the exchange bias interaction. It has been demonstrated that any stack sequence of LSMO on BFO\,\cite{Wu:13} or BFO on LSMO\,\cite{Yu:10, Yu:12, Xu:14} can exhibit an exchange bias interaction at temperatures below 100\,K as the blocking temperature\,\cite{Yu:10}. The exchange bias is attributed to the uncompensated Fe moments at the interface based on the observation of ferromagnetic Fe atoms registered by X-ray magnetic circular dichroism signals\,\cite{Yu:10}. However, recent polarized neutron reflectivity (PNR) and X-ray resonant magnetic reflectometry measurements revealed no ferromagnetic Fe at the interface due to the formation of a magnetically diluted interface\,\cite{Bertinshaw:14}. In contrast, another report of PNR measurement on (LSMO)$_6$-(BFO)$_5$ superlattices shows ferromagnetic Fe within the BFO layers\,\cite{Singh:14}. Moreover, some theoretical models have been suggested to describe the exchange bias coupling at the interface of LSMO and BFO. A microscopic model considering magnetic exchange interaction and ferroelectric Coulomb interaction confirms the induced magnetization in BFO upon charge transfer\,\cite{calderon:11}. On the other hand, density functional calculations predict the formation of uncompensated Fe moments not at an abrupt interface of BFO and LSMO, but rather at an intermixed and rough one\,\cite{Neumann:12}. 

In this paper, we study two different stack sequences, namely STO/BFO/LSMO (SBL) and STO/LSMO/BFO (SLB) in which the STO stands for SrTiO$_3$ substrate. We demonstrate the role of interface roughness on the exchange bias at the interface. Our findings show no exchange bias at the sharp interfaces, but a sizable one at rough and chemically intermixed interfaces. Furthermore, we use the temperature dependence of coercivity and exchange bias field to analyze the spin coupling at the interface of BFO and LSMO. 
	
Multistack heterostructures of SBL and SLB were grown on STO (001) single crystalline substrate using pulsed laser deposition. Magnetoresistance (MR) measurements were performed on Hall-bar like microstructures which were fabricated using electron beam lithography and Ar$^+$ ion milling (for details on sample fabrication, see the supplementary information\,\cite{S1}).
\begin{figure}[htbp!]
\centering
\includegraphics[width=1\columnwidth,clip=true]{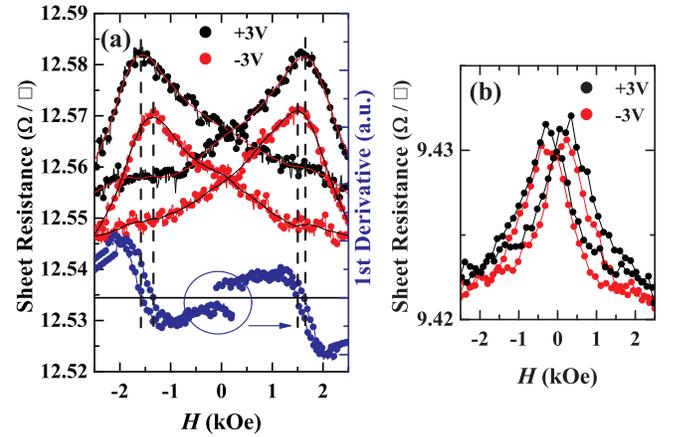}
\caption{MR curves measured at 10\,K after poling the BFO with a $\pm3$\,V pulsed gate voltage for (a) STO/SRO/BFO/LSMO and (b) STO/LSMO/BFO/SRO heterostructures in which the SRO (SrRuO$_3$) was used as an electrode. The solid lines in (a) are the smoothed MR curves and the blue curves are their first derivatives. The thicknesses of SRO, BFO and LSMO layers were fixed to 30, 25 and 7\,nm, respectively. The curves in (a) show a change in sheet resistance of LSMO, $H_{C}$ and $H_{Eb}$. For (b), however, no significant change is registered after poling with different polarities.}\label{Fig:MR}
\end{figure}

First, we present the MR curves of the SBL heterostructure showing, respectively, a coercive field ($H_{C}$) and exchange bias field ($H_{Eb}$) of 1600$\pm40$ and 20$\pm20$\,Oe after poling with +3\,V and 1400$\pm40$ and 50$\pm20$\,Oe after poling with -3\,V gate voltage (Fig.~\ref{Fig:MR}a). In order to evaluate the coercive and exchange bias fields from MR curves, first a Savitzky-Golay smoothing method with a 20 point window (polynomial order of 2) was applied to the MR curves, then the first derivatives of the smoothed curves were used to evaluate the maxima reliably as shown in Fig.~\ref{Fig:MR}a.
The $H_{C}$ and $H_{Eb}$ values were evaluated using the following equations\,\cite{Radu:08}:
\begin{equation} \label{eq:1}
{H_C} = \frac{\text{$H^{+}$\,-\,$H^{-}$}}{\text{2}} , {H_{Eb}} = \frac{\text{$H^{+}$\,+\,$H^{-}$}}{\text{2}} \end{equation} 
where $H^{+}$ and $H^{-}$ are the magnetic fields with positive and negative polarities in which the MR curves are at their maximum. After applying a positive gate voltage, the structure exhibits a high resistance state, high coercive field and low exchange bias field, while switching the polarization of ferroelectric BFO brings the system to a low resistance state, low coercive field and high exchange bias field. These results are in agreement with previously reported data\,\cite{Wu:13} which was however measured on SLB heterostructures suggesting that applying a positive gate voltage leads to pinning of more Fe moments to the spin structure of the LSMO\,\cite{Wu:13}. On the other hand, the increase in resistivity upon applying a positive gate voltage suggests that pinned Fe moments act as scattering centers. We characterized our SLB heterostructure as well, and found that the MR curves after applying gate voltages with different polarities show a small change in $H_{C}$ and $H_{Eb}$, and almost no change in the sheet resistivity (see Fig.~\ref{Fig:MR}b). Comparing the MR curves of the two stacks, one notices that the SLB heterostructure exhibits lower coercive field and sheet resistance than the SBL heterostructure, indicating a more conductive LSMO associated with less pinned Fe moments at the interface in the SLB heterostructure in agreement with the tendency observed for SBL heterostructure. 

\begin{figure}[htbp!]
\centering
\includegraphics[width=1.0\columnwidth,clip=true]{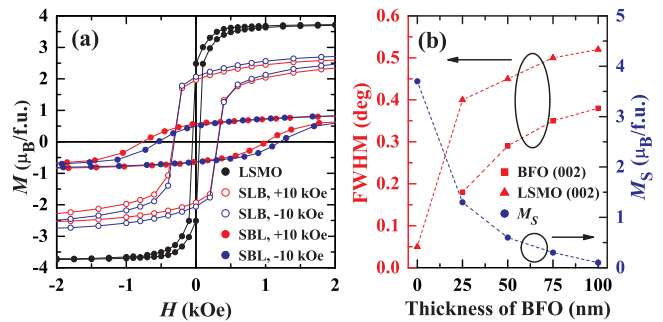}
\caption{(a) Magnetic field-dependent magnetization measurement for STO/BFO/LSMO (SBL), STO/LSMO/BFO (SLB) heterostructures and a single layer LSMO thin film (50\,nm). The thicknesses of the LSMO and BFO in the heterostructures are 7 and 50\,nm, respectively. The hysteresis loops were recorded at 10\,K after field cooling under applied magnetic field of $\pm$10\,kOe. (b) The effect of the BFO thickness on the saturation magnetization ($M_S$) of LSMO; and on the Full Width at the Half Maximum (FWHM) of rocking curve measurements on (002) reflections of LSMO and BFO.}\label{Fig:SQUID}
\end{figure}

In order to understand the results from MR measurements, the heterostructures were further characterized by SQUID after field cooling from 380\,K under applied magnetic fields of $\pm$10\,kOe. In Fig.~\ref{Fig:SQUID}a, the hysteresis loops of the SBL and the SLB heterostructures measured at 5\,K are shown. Both heterostructures show a higher coercive field with respect to the single LSMO layer that is an indication of the existence of the uncompensated moments (either pinned to ferromagnetic or antiferromagnetic layers) at the interface\,\cite{Bruck:08}. While the SBL heterostructure shows a clear exchange bias shift ${i.e.}$ shift in the hysteresis loop from zero applied magnetic field, the SLB shows no such effect. 

The $H_{C}$ and $H_{Eb}$ for the SBL heterostructure upon field cooling under +10\,kOe field are evaluated as 870 and 140\,Oe using equation \ref{eq:1}. Here, $H^{+}$ and $H^{-}$ represent the corresponding coercive fields upon field cooling under magnetic field with positive and negative polarities, respectively. Upon field cooling under -10\,kOe field, a positive shift of 200\,Oe leads to $H_{Eb}$ of 340\,Oe. Comparing the MR curves and hysteresis loops of both stack sequences (Fig.~\ref{Fig:MR}a,b and Fig.~\ref{Fig:SQUID}a), it is obvious that the SBL heterostructure exhibits a higher $H_{C}$ and exchange bias can only be detected in this stack sequence. We observed the same $H_{C}$ and $H_{Eb}$ in the MR curves of $\pm$10\,kOe field cooled SBL heterostructures (thicknesses of BFO and LSMO were 50 and 7\,nm, respectively)\,\cite{S1}.

Moreover, concerning the hysteresis loops, the saturation magnetization ($M_S$) decreases from 3.7\,$\mu_B$/f.u. for the single LSMO layer to approximately 2.6 and 0.8\,$\mu_B$/f.u. for the SLB and SBL heterostructures, respectively. The reason is likely to be due to the crystal quality of the LSMO layer. Rocking curve measurements on LSMO (002) and BFO (002) reflections (not shown here) reveal a full width at the half maximum (FWHM) of 0.08$^{\circ}$ and 0.21$^{\circ}$ for the SLB and 0.29$^{\circ}$ and 0.45$^{\circ}$ for the SBL, respectively. This means that, the amount of defects in the LSMO within the SLB heterostructure is lower than the one within the SBL giving a higher volume fraction with long range magnetic order contributing to the saturation magnetization. Note that the magnetization is normalized to the volume of a formula unit. The lower crystal quality of BFO and LSMO in SBL heterostructure is due to the fact that the growth of BFO is accompanied by the formation of defects and dislocations leading to the formation of a rough BFO surface on which the LSMO is then grown. This might be related to the fact that LSMO and STO are true perovskite structures, while BFO has a rhombohedral crystal symmetry which is only approximated by a perovskite structure. So different crystallographic variants will coexist on a cubic single crystalline substrate surface, while LSMO can grow truly epitaxially on STO. Such tendency of reduced $M_S$ was also observed in a series of SBL heterostructures having different thicknesses of BFO. As shown in Fig.~\ref{Fig:SQUID}b, the FWHM of the rocking curve of the (002) reflection of both BFO and LSMO layers increases with the thickness of BFO, while the $M_S$ decreases. Therefore, these results signify that the reduction of $M_S$ is a function of effective fraction of volume owing the magnetic long-range order.

\begin{figure}[htbp!]
\centering
\includegraphics[width=1.0\columnwidth,clip=true]{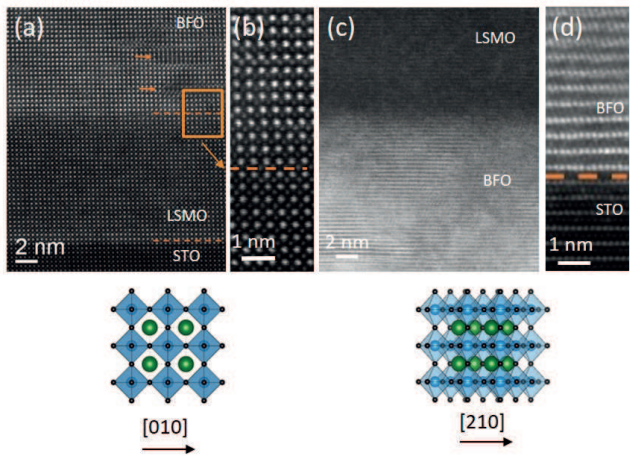}
\caption{HAADF-STEM images of (a) the STO/LSMO/BFO (SLB) and (c) the STO/BFO/LSMO (SBL) heterostructures. (b) High-resolution image of LSMO/BFO interface marked with orange square in (a). (d) High-resolution image of a rather sharp STO-BFO interface for the sample shown in (c). Orange dashed lines mark the interfaces. The thicknesses of BFO and LSMO for both heterostructures are 25 and 7\,nm. Note that the in-plane cut of the STO substrate for SLB and SBL are [010] and [210], respectively, leading to different interatomic distances. The schematic views of perovskite structure with respect to the corresponding in-plane cut are illustrated in which the green, blue and black spheres depict (La/Sr, Bi), (Mn, Fe) and O, respectively.}\label{Fig:TEM}
\end{figure}

To examine the possible correlation between the exchange bias interaction and the atomic structure at the interface, HAADF-STEM was conducted on both SBL and SLB heterostructures shown in Fig.~\ref{Fig:TEM}\,\cite{S1}. For the SLB, both LSMO/BFO and LSMO/STO interfaces are sharp with no dislocations or other defects (Fig.~\ref{Fig:TEM}a). However, some stacking faults and low-angle grain boundaries are visible inside the BFO film matrix which are marked with arrows in Fig.~\ref{Fig:TEM}a. The pseudo-cubic lattice structures of both perovskite layers for the SLB heterostructure are revealed in the image shown in Fig.~\ref{Fig:TEM}b. Our magnetometry and MR measurement (see Fig.~\ref{Fig:SQUID}a and Fig.~\ref{Fig:MR}b) show no exchange bias at this sharp interface of LSMO-BFO, even though for this sequence, exchange bias was previously reported in Ref.\,\cite{Wu:13}. Unlike the SLB, the SBL heterostructure shows a rough BFO/LSMO interface (Fig.~\ref{Fig:TEM}c), yet a rather sharp STO/BFO interface (Fig.~\ref{Fig:TEM}d). The high-resolution TEM images and FFT patterns (not shown here) of this sample reveal that both BFO and LSMO layers have relatively poor crystalline quality for this stack sequence.


While, EDX line scans reveal no significant intermixing of atomic species at the interface for the SLB heterostructure, an approximately 1-2\,nm intermixed region was found for the SBL heterostructure. For such a rough interface, we actually observe sizable exchange bias effect (see Fig.~\ref{Fig:SQUID}a and Fig.~\ref{Fig:MR}a) whose origin cannot be explained by Mn-O-Fe hybridization through orbital reconstruction at the interface as suggested in Ref.\,\cite{Yu:10} since such scenario requires a sharp interface. The observed exchange bias can be explained rather with a model suggested in Ref.\,\cite{Neumann:12}. In this model, an intermixed interface which exhibits exchange interaction is energetically favored. Such an interface is associated with uncompensated Fe moments when the BFO at the interface is positively charged. Moreover, compared to a sharp interface, the difference between positively and negatively charged BFO at such a rough interface is drastically smaller. This can also be confirmed by comparing the exchange bias and coercive fields differences at two different polarities of our heterostructures and the ones reported in Ref.\,\cite{Wu:13}. 

\begin{figure}[htbp!]
\centering
\includegraphics[width=1\columnwidth,clip=true]{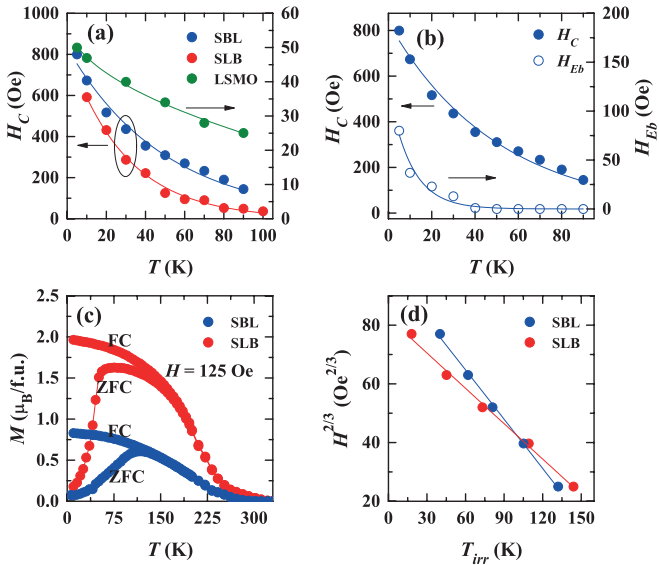}
\caption{(a) Temperature dependence of coercivity for the STO/BFO/LSMO (SBL) and STO/LSMO/BFO (SLB) heterostructures and a 50\,nm LSMO film. (b) Temperature dependence of coercive and exchange bias field of the SBL heterostructure shown in (a). (c) ZFC and FC magnetization curves of the SBL and SLB heterostructures measured under applied magnetic field of 125\,Oe. (d). Irreversible magnetization temperature evaluated from the ZFC-FC measurements under different applied magnetic fields. The solid lines in a, b, and d are fitted lines. The thicknesses of BFO and LSMO for both heterostructures are 25 and 7\,nm.}\label{Fig:SG}
\end{figure}

In order to understand the temperature-dependent pinning behavior of Fe moments in the SBL and SLB heterostructures, the temperature-dependence of the coercivity, $H_{C}$($T$), needs to be ascertained by measuring the hysteresis loops at different temperatures. The coercivity of single domain particles decreases with temperature due to the thermal fluctuations following the $H_C$\,$\propto$\,$T^{-1/2}$ relation\,\cite{Cullity:72}, which is the case here for the single layer LSMO film shown in Fig.~\ref{Fig:SG}a. However, such dependence cannot be found for neither the SBL nor the SLB heterostructures (Fig.~\ref{Fig:SG}a). Instead the $H_{C}$($T$) measurements follow the following phenomenological functional dependence:
\begin{equation} \label{eq:2}
{H_C(T)} = {H_C(0)\,\exp\,(-T/T_0)} \end{equation}
in which $H_C(0)$ is the coercive field at 0\,K and $T_0$ is a constant. Such an exponential decay of the coercivity with temperature has been observed in the magnetically frustrated systems in which the competing magnetic domains form a spin-glass state\,\cite{Moutis:01, Karmakar:08, Ding:13, Huang:08}. Furthermore, we find the same exponential decay for $H_{Eb}$ with temperature in the SBL heterostructure as shown in Fig.~\ref{Fig:SG}b. 


In addition, the zero-field cooled (ZFC) and field cooled (FC) temperature-dependent magnetization measurements\,\cite{S1} for both heterostructure sequences exhibit an irreversible behavior as the ZFC and FC curves separate at $T<T_{irr}$ in which $T_{irr}$ is the irreversible magnetization temperature (Fig.~\ref{Fig:SG}c). The SLB heterostructure exhibits a higher magnetization owing to its higher crystal quality as discussed before (Fig.~\ref{Fig:SQUID}a). Such irreversible behavior has not been observed for 7\,nm single layer LSMO films\,\cite{S1}.
Thouless and de~Almeida\,\cite{Almeida:78} defined the boundaries of a spin-glass state as a function of temperature and applied magnetic field based on the $Sherrington-Kirkpatrick$ Model\,\cite{Sherrington:75}, a mean field model which has been widely used to describe spin-glass systems. Thouless and de~Almeida showed that in a spin-glass system, there is a linear dependence of $T_{irr}$ on $H^{2/3}$. We find that the $T_{irr}$ decreases with applied magnetic field for ZFC and FC measurements and as shown in Fig.~\ref{Fig:SG}d, for both heterostructures, there is a linear dependence between $T_{irr}$ and $H^{2/3}$ following the so called de Almeida-Touless line\,\cite{Almeida:78}:
\begin{equation} \label{eq:3}
{H(T_{irr})/\Delta J} \propto {(1- T_{irr}/T_F)^{3/2}} \end{equation}
where $H(T_{irr})$ is the applied magnetic field under which the ZFC and FC measurements were performed. $\Delta$$J$ is the width of the distribution of exchange interaction and $T_F$ is the zero-field spin-glass freezing temperature which is found to be 176 and 206\,K for the SBL and SLB heterostructures, respectively. The temperature dependence of coercivity and the field-dependence of $T_{irr}$ suggest a spin-glass-like state at the interface of BFO and LSMO independent of the stack sequence and therefore the roughness of the interface. We speculate that the observation of a spin-glass-like behavior is indicative to result from the complex spin structure of the BFO. In fact, other studies suggest a spin-glass behavior for amorphous\,\cite{Nakamura:93} and (111)-oriented single crystalline thin films\,\cite{Singh:09} of BFO. Due to the lack of any sizable net magnetization for our single BFO films, we could not investigate such a scenario using magnetometry measurements. Furthermore, the SLB heterostructure $i.e.$ the heterostructure with the sharp interface exhibits a higher $T_F$ in comparison to the SBL heterostructure $i.e.$ the heterostructure with the relatively rough interface, indicating a more persistent spin-glass-state for the sharp interfaces. This along with the fact that we observe the possible spin-glass-like behavior for both heterostructures suggest that the formation of a spin-glass-like state seems to be independent from the roughness of the interfaces. Obviously, the occurrence of a spin-glass-like state is not directly linked to the observation of exchange bias at the rough interface of SBL heterostructure, and rather reveals the complexity of the spin structure at the interface.

In conclusion, we have investigated the exchange bias interaction at sharp and rough interfaces of LSMO and BFO using MR and SQUID measurements. While the rough interface reveals an exchange bias effect, the sharp interface does not show any sizable exchange bias. Nevertheless, both stack sequences exhibit higher coercive fields compared to a single LSMO layer which is an indication of existence of uncompensated Fe moments at the interface that are either pinned to ferromagnetic LSMO or antiferromagnetic BFO. Moreover, both sequence stacks show a spin-glass-like behavior as the temperature-dependent coercivity and the field-dependent irreversible magnetization temperature curves suggest. While the spin-glass-like behavior seems to be independent of the interface roughness, it reveals the complex spin structure which may be associated with the interface of LSMO and BFO. Our results show that one needs to carefully determine not only the coupling but also the structure before such system is considered to be used as a future electric field controlled device. Our results can be explained well by density functional calculation reported in Ref.\,\cite{Neumann:12} predicting an exchange bias effect at rough interfaces. Based on these DFT calculation, in abrupt interfaces (SLB heterostructures), no exchange coupling occurs since the Fe-Mn exchange energy has to be 8 times more than the one of Fe-Fe. Furthermore, the Fe-Mn exchange coupling has to be antiferromagnetic in this case. In a rough and atomically intermixed interface (SBL heterostructures), however, only the antiferromagnetic Fe-Mn exchange coupling has to occur. Therefore, the suggested orbital reconstruction model at the interface of LSMO and BFO\,\cite{Wu:13, Yu:10, calderon:11} cannot explain our results, while it has been used to describe the exchange bias effect in similar stacks. This could be due to the complex magnetic structure of BFO since previous studies suggest a cycloidal model of spin ordering along $<110>$ direction\,\cite{Lebeugle:08}, a spin-glass-like state\,\cite{Singh:09}, and even weak ferromagnetism at low temperatures\,\cite{Han:13}. Therefore, the samples across different studies fabricated by different groups need to be compared carefully and have to be individually characterized. Since there is no univocal theoretical picture and given our surprising findings, our result show that structural characterization is the key for understanding the measured electrical properties. In addition, for applications, the fatigue poses another challenge that needs to be addressed and studied in these systems\,\cite{S1}.  

The authors would like to acknowledge the EU projects (IFOX, NMP3-LA-2010 246102 and Inspin, FP7-ICT-2013-X 612759) and the Graduate School of Excellence MAINZ (GSC 266 MAINZ) as the funding agents.

\end{document}